# Blockchain-Enabled Artificial Intelligence and AI Agents for Secure Data Sharing and Cybersecurity Applications

**Harsh Verma**[1]

*Palo Alto Networks, Artificial Intelligence, United States*

**Abstract:** Blockchain and artificial intelligence (AI) are converging into a single infrastructural layer for securing data sharing, model integrity, and autonomous decision-making across distributed systems. This paper presents a meta-synthesis that draws together four constituent studies covering adversarial machine learning, AI-powered anomaly detection in cloud environments, automated vulnerability patching by multi-agent large language model (LLM) pipelines, and the broader landscape of securing AI systems across their lifecycle and situates their findings within the emerging literature on blockchain-enabled AI and autonomous AI agents. Each constituent study addresses a distinct point of failure in modern AI-driven security operations: the integrity of training data and model behavior, the reliability of real-time monitoring, and the trustworthiness of automated code remediation. We argue that blockchain's properties of immutability, decentralized consensus, and verifiable provenance directly address a gap common to all three: the difficulty of establishing trust in data, models, and autonomous agents that operate without a central authority. Building on real-world research on blockchain-secured data sharing, federated learning, and multi-agent coordination, we propose a layered reference architecture that couples adversarially hardened models, blockchain-anchored data provenance, AI-driven anomaly detection, and smart-contract-governed multi-agent remediation. We conclude by identifying open problems in scalability, privacy-transparency trade-offs, and the governance of autonomous agents that must be resolved before such integrated systems can be trusted in production-critical environments.

## 1. Introduction

Artificial intelligence has moved from a peripheral analytics tool to a core operational component of cybersecurity, finance, healthcare, and critical infrastructure. As this shift has occurred, four distinct but interlocking problems have emerged in parallel research streams: how to keep machine learning models themselves from being manipulated by adversaries, how to detect anomalous behavior across sprawling, ephemeral cloud infrastructure, how to remediate vulnerabilities in code including decades-old legacy systems faster than attackers can exploit them, and how to govern the growing population of autonomous AI agents that increasingly make or execute these decisions with limited human oversight. This paper synthesizes four prior studies that each address one of these problems in isolation, and asks a further question that none of them fully answers: what happens when the data, models, and agents themselves cannot be trusted to have been unmodified in the first place?

Blockchain technology, defined broadly as an append-only, cryptographically linked, and distributedly maintained ledger, has been proposed repeatedly in the literature as an answer to exactly this trust problem. Distributed ledgers provide tamper-evident logging, verifiable provenance for data and model artifacts, and, increasingly, a coordination substrate for autonomous software agents that must act without a central intermediary. Recent surveys of blockchain-enabled AI describe the pairing as mutually reinforcing: AI supplies the analytical and predictive capability that blockchains lack natively, while blockchain supplies the transparency, traceability, and secure sharing that AI systems, particularly those trained on distributed or sensitive data, structurally lack (Dilawar, Wang, Xin, & Yun, 2024). At the same time, a fast-growing body of work on AI agents operating within decentralized systems has begun to treat blockchain not merely as a storage layer but as a trust and governance layer for multi-agent collaboration (Karim, 2025; Qi, Zhu, Zhang, Li, & Zhou, 2025).

This paper is organized as follows. Section 2 reviews the background of blockchain and AI convergence relevant to security applications. Section 3 synthesizes the four constituent studies, treating them as case studies covering the model, infrastructure, and code layers of AI-driven security operations. Section 4 examines blockchain specifically as a trust substrate for AI data and models. Section 5 turns to AI agents operating on or alongside blockchains, extending the multi-agent remediation pipelines discussed in Section 3 into a decentralized, verifiable setting. Section 6 proposes an integrated layered architecture. Section 7 discusses open challenges, and Section 8 concludes.

## 2. Background: The Convergence of Blockchain and AI in Cybersecurity

### 2.1 Blockchain Fundamentals Relevant to Security

A blockchain is a distributed ledger in which transactions are grouped into cryptographically chained blocks and replicated across a peer-to-peer network, with agreement on the ledger's state reached through a consensus protocol rather than a single trusted party. For security applications, three properties matter most: immutability, which makes retroactive tampering with recorded data computationally infeasible; decentralization, which removes single points of failure and control; and programmability through smart contracts, which allows rules such as access control policies or validation gates to be enforced automatically and transparently. Surveys of blockchain-enabled data sharing have repeatedly used these properties to address authentication and access-control gaps in Internet of Things (IoT) and cloud environments, for instance through multi-authority, attribute-based searchable encryption schemes that let data owners retain fine-grained control over who can query their data even after it is shared (Zhang, Xia, Gao, Ma, & Chen, 2025).

### 2.2 AI in Cybersecurity: A Rapidly Maturing but Fragmented Field

AI's role in cybersecurity has been reviewed extensively over the past decade, from early framings of machine learning as a defensive tool against intrusion (Li, 2018) to comprehensive surveys evaluating the efficiency of AI and ML techniques across the full range of security solutions, including intrusion detection, malware classification, and fraud prevention (Ozkan-Okay et al., 2024). More recent reviews emphasize that AI and cybersecurity have become mutually entangled: AI both defends systems and, when embedded insecurely, becomes itself a novel attack surface that traditional security tooling was never designed to monitor (Paramesha, Rane, & Rane, 2024; Mohamed, 2025). This entanglement is the throughline that connects the four constituent studies synthesized in Section 3, and it is precisely the layer at which blockchain-based trust mechanisms have been proposed as a complementary safeguard.

### 2.3 Why Couple Blockchain with AI?

The rationale for combining the two technologies is symmetrical. AI systems depend on large volumes of training and telemetry data whose provenance, ownership, and integrity are often unverifiable once the data has passed through multiple intermediaries; blockchain's tamper-evident ledger offers a way to record and later verify that lineage. Conversely, blockchains generate enormous volumes of transactional and behavioral data that benefit from AI-driven analysis, and blockchain networks themselves, particularly their consensus and smart-contract layers, can be optimized using AI (Dilawar et al., 2024). This bidirectional relationship has been formalized under the umbrella of "blockchain-based trustworthy AI," which surveys the technology's use across the AI development lifecycle to promote transparency, fairness, robustness, and accountability in systems that would otherwise remain opaque "black boxes" (Guo, Yu, Guo, & Xiang, 2023). Industry-facing analyses reach a similar conclusion from a different angle, reporting that data privacy and security protocols are, by a wide margin, the factor executives cite most often as driving further blockchain-AI integration (Kshetri, 2025).

## 3. Synthesis of Prior Work: Four Facets of the AI Security Lifecycle

The four constituent studies underlying this meta-synthesis were not originally framed around blockchain, but each addresses a layer of the AI security lifecycle at which a verifiable, decentralized trust mechanism would materially change the threat model. This section summarizes each study's core contribution and identifies that latent connection.

### 3.1 Securing the Model Itself: Adversarial Threats Across the ML Lifecycle

Two of the constituent studies, Secure AI Systems: Protecting Machine Learning Models from Emerging Cyber Threats (Verma, 2026) and Adversarial Machine Learning: Security Risks and Defense Strategies in AI-Driven Applications (Verma, 2025), together map the attack surface of machine learning models across the data, model, and deployment layers. Both identify data poisoning, adversarial evasion at inference time, backdoor insertion, model extraction, and model-inversion privacy attacks as the dominant threat categories, echoing taxonomies developed elsewhere in the adversarial machine learning literature (Olutimehin et al., 2025; Pelekis et al., 2025). On the defensive side, the studies converge on a similar set of countermeasures: adversarial training and robust optimization to harden models against evasion, defensive distillation, anomaly-based detection of malicious inputs, and privacy-preserving training techniques such as differential privacy, federated learning, and homomorphic encryption (Verma, 2025, 2026; Rigaki & Garcia, 2023).

Both studies stop short of asking how the integrity of the training data itself, or the provenance of a deployed model, can be verified independently of the organization that trained it. This is precisely the gap that blockchain-anchored data lineage and federated-learning coordination protocols are designed to close, a connection developed further in Section 4.

**3.2 Securing the Infrastructure: AI-Powered Anomaly Detection in Cloud Environments**

AI-Powered Anomaly Detection in Cloud-Based Applications (Verma, n.d.-a) addresses a different layer: the elastic, ephemeral infrastructure on which most modern AI systems, including the adversarially targeted models discussed above, are deployed. The study argues that static, threshold-based monitoring cannot keep pace with containerized and microservice-based workloads that scale continuously and redeploy frequently, and proposes a framework combining supervised learning, unsupervised clustering, and deep architectures such as autoencoders and recurrent neural networks to build adaptive behavioral baselines from log, metric, and network-trace telemetry. In an evaluation exceeding 500,000 records across five benchmarked model families, hybrid ensemble approaches produced the strongest detection performance, and the study links this improvement to faster incident response and reduced alert fatigue for on-call security teams.

This anomaly-detection layer is a natural point of integration with blockchain-based audit logging: once an AI monitoring pipeline flags a deviation, recording that alert, the evidence behind it, and any subsequent remediation action on an immutable ledger converts a purely internal detection event into something independently auditable by regulators, customers, or third-party security assessors, a capability increasingly demanded of AI-driven monitoring in regulated industries (Tabassi, 2023).

**3.3 Securing the Code: Multi-Agent LLM Pipelines for Vulnerability Patching**

Automated Vulnerability Patching in Legacy Code Using LLMs and Multi-AI Agents (Verma, n.d.-b) traces the methodological evolution of automated program repair from template- and heuristic-based systems (Liu, Koyuncu, Kim, & Bissyandé, 2019) through fine-tuned transformer models, zero-shot and conversational prompting of general-purpose LLMs (Pearce, Tan, Ahmad, Karri, & Dolan-Gavitt, 2023; Xia & Zhang, 2024; Ahmed & Devanbu, 2023), and finally to multi-agent pipelines in which specialized agents divide the labor of vulnerability detection, localization, patch generation, and validation (Zhang, Ruan, Fan, & Roychoudhury, 2024; Tao, Zhou, Zhang, & Cheng, 2024; Lee et al., 2024). This architectural decomposition mirrors the broader shift toward multi-agent collaboration described in the general LLM-agent literature (Talebirad & Nadiri, 2023), and the report finds it particularly well-suited to legacy code, where sparse documentation and thin automated test coverage make single-shot patch generation unreliable, since different agents can be assigned to reconstruct missing context, cross-reference historical commits, or synthesize missing test oracles.

The report's central caution is that this literature has been validated almost exclusively on actively maintained, well-organized open-source repositories, and that the gap between these benchmarks and the scale and disorganization of real enterprise legacy systems remains empirically unclosed. It further identifies trust and explainability mechanisms for security-critical maintenance work as the field's most consequential open problem a problem for which verifiable, on-chain validation of each agent's patch and each reviewer's sign-off would provide a natural, auditable answer.

**3.4 A Common Thread: Trust Without a Central Authority**

Read together, these three lines of work describe a full pipeline: hardening the model against manipulation, monitoring the infrastructure that hosts it, and autonomously remediating the vulnerabilities discovered in its supporting code. What none of them can, on their own, guarantee is that the data feeding the anomaly detector, the model being hardened, or the patch produced by an autonomous agent has not itself been tampered with somewhere along a multi-party pipeline. This is the trust problem that blockchain-enabled AI research has been built to address, and it is the subject of Section 4.

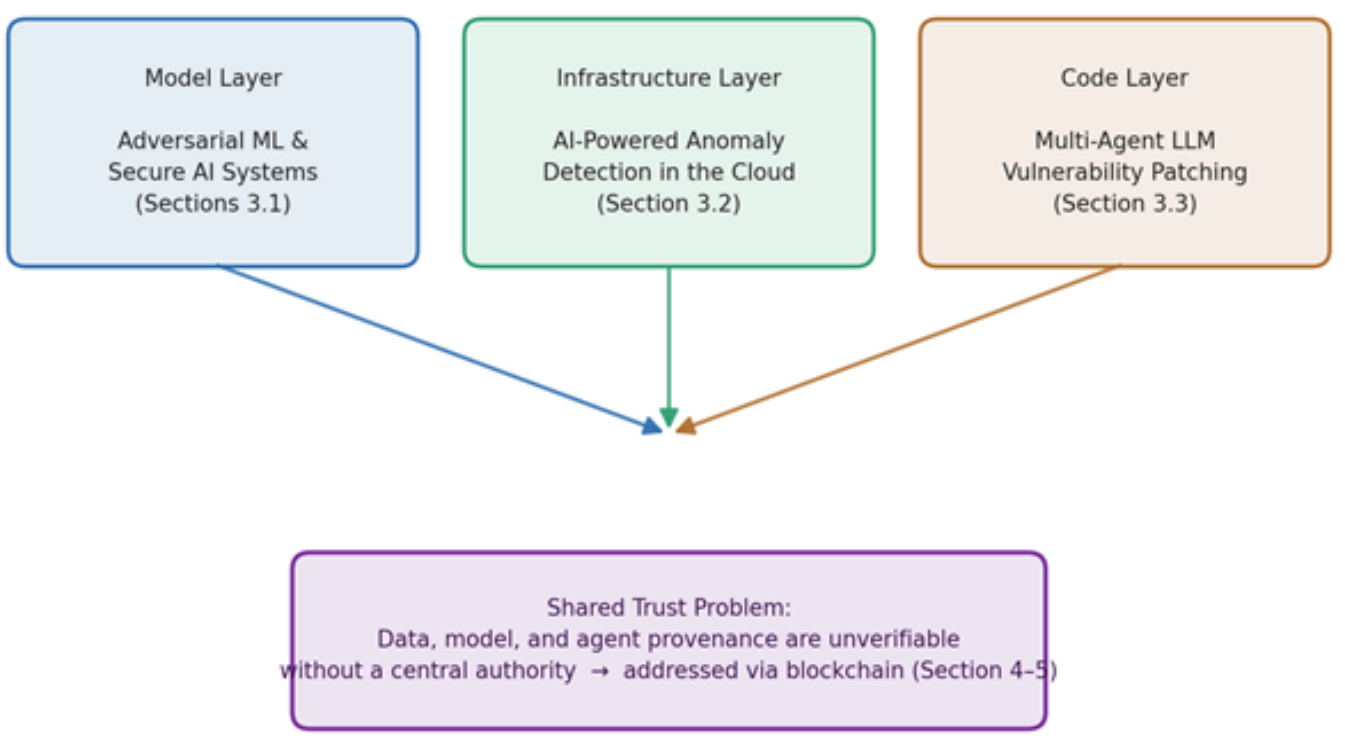

***Figure 1*** *Three facets of the AI security lifecycle model, infrastructure, and code converge on a common trust gap that blockchain-based mechanisms are used to close (Sections 4-5).*

## 4. Blockchain as a Trust Substrate for AI Data, Models, and Monitoring

### 4.1 Provenance and Integrity of Training and Telemetry Data

A foundational strand of blockchain-enabled AI research treats the ledger as a way to establish real, verifiable big data for AI consumption in environments where data is scattered across mutually distrusting stakeholders. The SecNet architecture, for example, combines blockchain-based data sharing with ownership guarantees, an AI-based secure computing platform, and blockchain-optimized security rules to enable trusted data sharing at internet scale, explicitly aiming to give AI systems a more reliable data source while using AI to strengthen the blockchain layer in return. Multi-authority, attribute-based searchable encryption schemes extend this idea to IoT-scale data sharing, decentralizing the authorization process itself so that no single authority becomes a point of compromise or failure (Zhang et al., 2025). Earlier IoT security work by one of the constituent studies' own author group similarly emphasized the need for secure, real-time management of heterogeneous IoT data streams as a prerequisite for any downstream analytics (Islam, Verma, Khan, & Kantarcioglu, 2019), a concern that anticipates the provenance problem blockchain-based data-sharing schemes now address more directly.

### 4.2 Blockchain-Secured Federated Learning

The privacy-preserving training techniques discussed in the adversarial machine learning literature, particularly federated learning, gain an additional layer of assurance when coordinated through a blockchain rather than a single aggregating server. Blockchain-based federated learning schemes for industrial IoT settings replace the central aggregator, itself a potential single point of failure or manipulation, with a consortium of nodes that jointly verify model updates before they are incorporated into the global model, reducing the opportunity for a single compromised participant to poison the shared model undetected (Lu, Huang, Dai, Maharjan, & Zhang, 2019). Related work applies a similar pattern to privacy-preserving training on vertically partitioned datasets, combining a blockchain consortium with homomorphic encryption so that no trusted third party is required to coordinate multi-party model training (Kuo & Ohno-Machado, 2019, as surveyed in Wang, Feng, He, Tang, & Zhang, 2021).

### 4.3 Immutable Audit Trails for Detection and Remediation

For the anomaly-detection and vulnerability-patching pipelines described in Sections 3.2 and 3.3, blockchain's contribution is less about training-time trust and more about operational accountability. Recording each anomaly alert, its supporting telemetry hash, and the eventual remediation action, whether a human decision or an autonomous patch, on an append-only ledger converts internal security operations into an externally verifiable audit trail. Broader surveys of blockchain-based trustworthy AI frame this capability as a software development lifecycle concern in its own right, arguing that transparency and traceability must be engineered into AI systems from data collection through deployment and monitoring, not retrofitted after an incident (Guo et al., 2023). NIST's AI Risk Management Framework similarly treats traceability and accountability as core functions that organizations deploying AI systems, particularly in security-relevant contexts, are expected to demonstrate (Tabassi, 2023).

## 5. AI Agents on the Blockchain: Autonomous, Verifiable Collaboration

### 5.1 The Rise of Multi-Agent LLM Systems

The vulnerability-patching pipelines reviewed in Section 3.3 are one instance of a broader shift in applied AI toward decomposing complex tasks across cooperating, specialized agents rather than a single monolithic model (Talebirad & Nadiri, 2023). As these agents are increasingly deployed to take real, consequential actions, such as merging a security patch, executing a financial transaction, or coordinating a response to a detected intrusion, without direct human sign-off on each step, the question of how to establish trust between agents, and between agents and the humans who rely on them, becomes as pressing as the question of model accuracy itself.

### 5.2 Blockchain-Secured Multi-Agent Coordination

A rapidly growing literature addresses this question directly by using blockchain as the coordination and trust layer for multi-agent systems. A comprehensive survey on the interplay between AI agents and blockchain finds that agents can improve blockchain consensus efficiency and resource allocation, while blockchain in turn provides the secure, scalable collaboration substrate that decentralized multi-agent systems require, particularly for interoperability and privacy across organizational boundaries (Wang, Liu, & Xu, 2025). Complementary architectures formalize trust-aware communication protocols for decentralized multi-agent systems, using cryptographic primitives and on-chain operations to guarantee communication integrity, authenticity, and non-repudiation between autonomous agents that may never share a common administrative domain (Chaffer, Ballandies, & colleagues' line of work, as reviewed in recent decentralized multi-agent system architectures). Governance-focused designs go further still: one proposed architecture integrates agentic AI frameworks with Ethereum-based smart contract governance so that autonomous agents can self-regulate and execute tasks only under enforceable, on-chain rules, explicitly targeting the integrity, trust, and control challenges that arise once agents act without centralized oversight (Jackson, 2025). Incentive-focused designs such as DAO-Agent extend this further by using zero-knowledge proofs to verify each agent's contribution and allocate rewards fairly without exposing sensitive execution data on-chain (Xia & Xu, 2025).

### 5.3 Toward Smart-Contract-Governed Security Pipelines

Applied to the cybersecurity domain specifically, this pattern suggests a natural extension of the multi-agent patching pipelines reviewed in Section 3.3: detection, localization, patch-generation, and validation agents could register their outputs and pass validation gates through smart contracts rather than an internal orchestration layer alone, so that no single agent, or the infrastructure operator running it, can unilaterally approve a patch without a recorded, independently verifiable trail. Recent decentralized security architectures for industrial IoT already gesture toward this pattern, deploying autonomous AI agents at each edge gateway as a distributed detection layer, though current implementations caution that full blockchain consensus overhead remains impractical for the lowest-latency industrial control loops and is better suited to auditability and coordination than to real-time detection itself (Singh & Roy, 2026).

## 6. Toward an Integrated Architecture

Synthesizing Sections 3 through 5 suggests a layered reference architecture for blockchain-enabled AI in cybersecurity and secure data sharing, in which each layer directly answers a limitation identified in one of the four constituent studies:

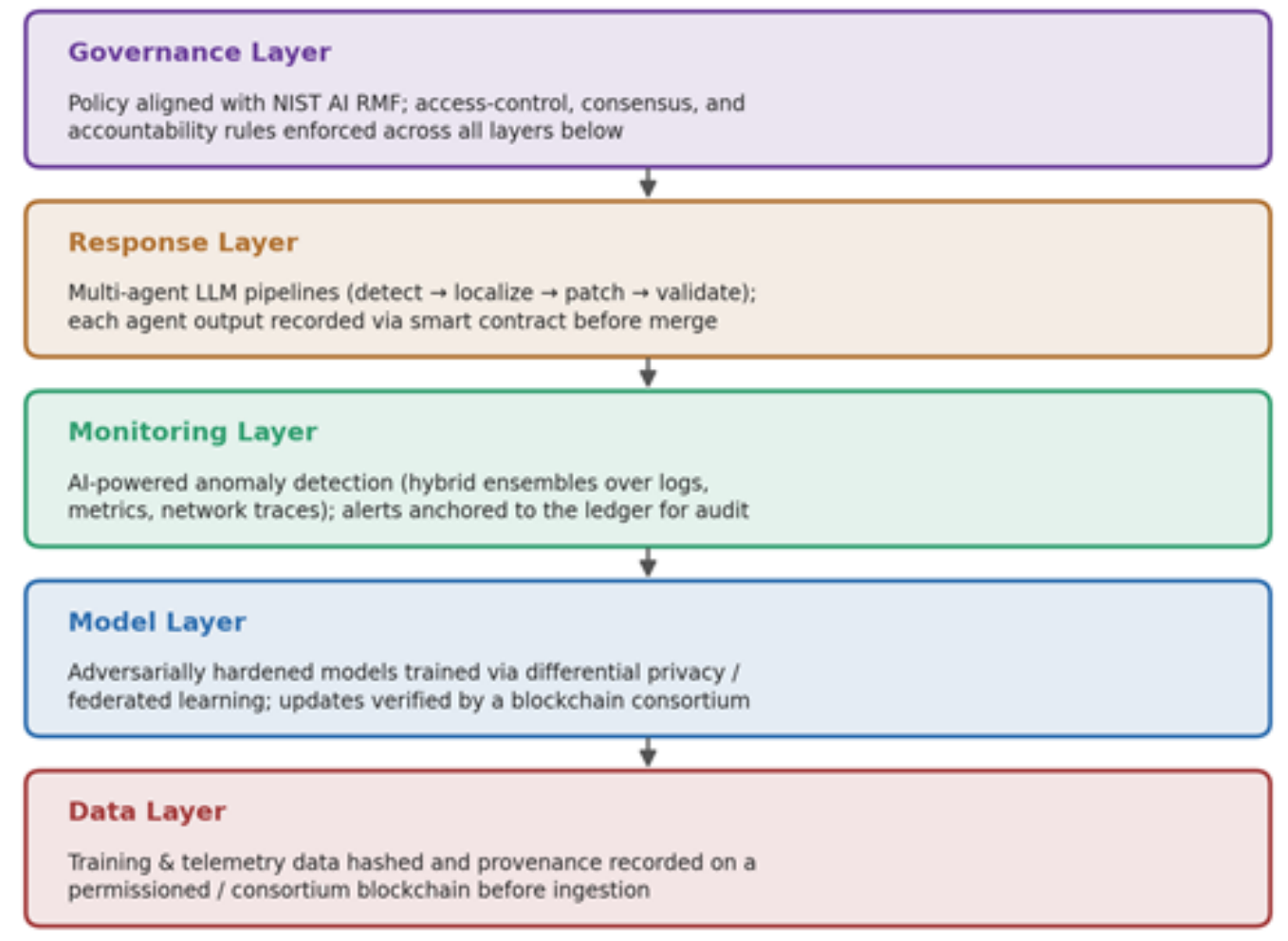

***Figure 2*** *Proposed layered reference architecture, in which each layer answers a limitation identified in one of the four constituent studies.*

**Data layer:** Training and telemetry data are hashed and their provenance recorded on a permissioned or consortium blockchain before ingestion, extending the trusted-data-sharing pattern of SecNet and attribute-based encryption schemes (Zhang et al., 2025) to close the provenance gap left open in Section 3.1.

**Model layer:** Models are trained using adversarially robust and privacy-preserving techniques adversarial training, differential privacy, federated learning with federated updates verified through a blockchain consortium rather than a single aggregator, following the pattern described in Section 4.2.

**Monitoring layer:** AI-powered anomaly detection, following the hybrid ensemble approach validated in Verma (n.d.-a), continuously profiles infrastructure behavior, with each triggered alert and its supporting evidence anchored to the ledger for later audit.

**Response layer:** Multi-agent LLM pipelines perform detection, localization, patch generation, and validation as described in Verma (n.d.-b), with each agent's output and each validation gate recorded via smart contract before a patch is merged, following the governance pattern of Jackson (2025) and Xia and Xu (2025).

**Governance layer:** Organization-level policy, aligned with frameworks such as NIST's AI Risk Management Framework (Tabassi, 2023), defines the access-control, consensus, and accountability rules enforced across the layers below.

No single constituent study, nor any single piece of the blockchain-AI literature reviewed here, implements this full stack; the contribution of this synthesis is to show that the pieces already exist in the literature and identify where they interlock.

## 7. Open Challenges and Research Directions

Several open problems must be resolved before an integrated architecture of this kind could be trusted in production-critical environments.

**Scalability versus latency.** Blockchain consensus introduces overhead that is often incompatible with the low-latency demands of real-time intrusion detection and industrial control loops; recent decentralized security architectures explicitly favor lightweight, non-blockchain consensus for detection while reserving blockchain for auditability and coordination (Singh & Roy, 2026).

**Privacy versus transparency.** The same immutability that enables auditability can conflict with the confidentiality of sensitive execution or training data; zero-knowledge and other privacy-preserving verification techniques are an active area of research aimed at resolving this tension without sacrificing either property (Xia & Xu, 2025; Rigaki & Garcia, 2023).

**The legacy-system validation gap.** As Verma (n.d.-b) documents, multi-agent patching research has been validated almost exclusively on well-organized open-source repositories, leaving open whether these methods, blockchain-secured or not, generalize to the scale and disorganization of real enterprise legacy systems.

**Standardization and benchmarking.** Verma (n.d.-a) similarly notes the absence of standardized datasets and evaluation frameworks for AI-powered anomaly detection, a gap that would need to be closed before blockchain-anchored audit trails could be compared meaningfully across organizations or vendors.

**Governance of autonomous agents.** As AI agents take on more consequential, less human-supervised roles within decentralized systems, questions of legal accountability, interoperability across heterogeneous agent frameworks, and collective governance mechanisms remain substantially unresolved (Kumar, Han, & colleagues' review of Web3 and AI agent integration challenges, 2025).

## 8. Conclusion

Blockchain-enabled artificial intelligence and autonomous AI agents are converging on a shared problem that runs beneath adversarial robustness, cloud anomaly detection, and automated vulnerability remediation alike: establishing trust in data, models, and agent behavior without relying on a single central authority. The four constituent studies synthesized in this paper each make substantial progress on one facet of AI-driven security, hardening models against adversarial manipulation, detecting anomalies across elastic cloud infrastructure, and automating vulnerability repair through cooperating LLM agents, but each also identifies a residual trust or auditability gap that blockchain-based mechanisms are increasingly being used to close elsewhere in the literature. Realizing an integrated, blockchain-secured AI security architecture will require sustained collaboration across the machine learning, distributed systems, and cybersecurity research communities, together with governance frameworks capable of holding increasingly autonomous systems accountable for the decisions they make.